\documentclass[sigconf]{acmart}

\renewcommand\footnotetextcopyrightpermission[1]{} 
\AtBeginDocument{%
  \providecommand\BibTeX{{%
    \normalfont B\kern-0.5em{\scshape i\kern-0.25em b}\kern-0.8em\TeX}}}

\input{mySymbol.sty}
\usepackage{subfigure}
\usepackage{graphicx}
\usepackage{xcolor}
\usepackage{color}

\begin{document}

\title{Transformer-Empowered Content-Aware Collaborative Filtering}


\author{Weizhe Lin}
\authornote{This is the work during Weizhe Lin's internship at Microsoft STCA.}
\affiliation{%
  \institution{University of Cambridge}
  \streetaddress{}
  \city{Cambridge}
  \country{United Kingdom}}
\email{wl356@cam.ac.uk}

\author{Linjun Shou}
\affiliation{%
  \institution{Microsoft STCA}
  \streetaddress{}
  \city{Beijing}
  \country{China}}
\email{lisho@microsoft.com}

\author{Ming Gong}
\affiliation{%
  \institution{Microsoft STCA}
  \streetaddress{}
  \city{Beijing}
  \country{China}}
\email{migon@microsoft.com}

\author{Pei Jian}
\affiliation{%
  \institution{Simon Fraser University}
  \streetaddress{}
  \city{British Columbia}
  \country{Canada}}
\email{jpei@cs.sfu.ca}

\author{Zhilin Wang}
\affiliation{%
  \institution{University of Washington}
  \streetaddress{}
  \city{Seattle}
  \country{United States}}
\email{zhilinw@uw.edu}

\author{Bill Byrne}
\affiliation{%
  \institution{University of Cambridge}
  \streetaddress{}
  \city{Cambridge}
  \country{United Kingdom}}
\email{bill.byrne@eng.cam.ac.uk}

\author{Daxin Jiang}
\authornote{The corresponding author.}
\affiliation{%
  \institution{Microsoft STCA}
  \streetaddress{}
  \city{Beijing}
  \country{China}}
\email{djiang@microsoft.com}
\renewcommand{\shortauthors}{Lin et al.}


\begin{abstract}
Knowledge graph (KG) based Collaborative Filtering is an effective approach to personalizing recommendation systems for relatively static domains such as movies and books, by leveraging structured information from KG to enrich both item and user representations. Motivated by the use of Transformers for understanding rich text in content-based filtering recommender systems, we propose Content-aware KG-enhanced Meta-preference Networks as a way to enhance collaborative filtering recommendation based on both structured information from KG as well as unstructured content features based on Transformer-empowered content-based filtering.
To achieve this, we employ a novel training scheme, Cross-System Contrastive Learning, to address the inconsistency of the two very different systems and propose a powerful collaborative filtering model and a variant of the well-known NRMS system within this modeling framework.
We also contribute to public domain resources through the creation of a large-scale movie-knowledge-graph dataset and an extension of the already public Amazon-Book dataset through incorporation of text descriptions crawled from external sources.
We present experimental results showing that enhancing collaborative filtering with Transformer-based features derived from content-based filtering outperforms strong baseline systems, improving the ability of knowledge-graph-based collaborative filtering systems to exploit item content information.

\end{abstract}




\maketitle

\section{Introduction}
\label{sec:introduction}


\textit{Collaborative Filtering} (\textbf{CF}) and \textit{Content-based Filtering} (\textbf{CBF}) are two leading recommendation techniques~\cite{takacs2009scalable}.
CBF models leverage only descriptive attributes of items (e.g. item description and category) and users (e.g. age and gender).
Users are characterized by the content information available in their browsing histories~\cite{thorat2015survey}.
CBF is particularly well-suited for news recommendation, where millions of new items are produced every day.
In contrast, CF systems study users' interactions in order to leverage inter-item, inter-user, or user-item dependencies in making recommendations.
The underlying notion is that users who interact with similar sets of items are likely to share preferences for other items.
CF systems are better suited for scenarios where the inventory of items grows slowly and where abundant user-item interactions are available.
Movie and book recommender systems are examples of such scenarios and will serve as the focus of this paper.

Collaborative filtering can benefit from the incorporation of knowledge graphs (KGs) to enrich the representations of users and items by drawing from external knowledge bases. 
KG-based CF models are particularly good at linking items to other related knowledge graph entities that serve as ``item properties'' and thus leveraging the structured content information in knowledge graphs.   
%

While KGs can readily incorporate structured content information (e.g. movie genre and actors),   unstructured content such as item descriptions, is largely unexploited.
For example, the two movies ``Interstellar'' and ``Inception'', have a very similar set of structured properties including genre, writer, and director, but their descriptions provide more fine-grained discriminative information, making it clear that one is about physics and universe and the other is about adventures and dreams.
Transformer-based models, such as BERT~\cite{devlin-etal-2019-bert} and GPT-2~\cite{radford2019language}, have shown great power in modeling descriptive content from natural language, and thus we investigate whether item/user representations can be enriched by more expressive representations derived from Transformers.

Recommender systems based on Transformers for Content-based Filtering have different properties from KG-based Collaborative Filtering systems, and fusing the two approaches is not yet addressed in the literature. 
The challenge in combining the two approaches mainly stems from the need to capture the co-occurrence of graph node features by graph convolution operations. 
This operation requires representations of graph nodes to be back-propagated and updated after each forward pass,
and thus it is prohibitively costly for large graphs where millions of nodes require transformer-generated embeddings.

This paper investigates computationally  efficient approaches to enrich CF models with both structured information derived from KGs and unstructured information derived from CBF. 
We introduce a framework that incorporates unstructured CBF features in CF training with efficient Cross-System Contrastive Learning that brings together the benefits of both structured and unstructured item properties.   We present two models: (1) KG-Enhanced Meta-Preference Network (KMPN),  a CF model that outperforms several baseline systems; and (2) a pre-trained variant of a transformer-based CBF model, NRMS~\cite{wu-etal-2019-neural-news}-BERT.
%


To research this problem, new datasets are needed for the study of these models, and should consist of (1) large-scale user-item interactions; (2) up-to-date, large-scale enormous knowledge graphs; and (3) rich descriptive attributes suitable for content-based movie and book recommendation.
%
Our contributions to research into these problems are (1) an extension of the widely-used Amazon-Book dataset through the addition of summary texts for books and (2) a new large-scale movie recommendation dataset.

In summary, our contributions are:
\begin{enumerate}
    \item We propose to merge unstructured content-based features into knowledge based CF through a simple but effective fusion framework based on Cross-System Contrastive Learning. 
In this, a powerful KG-CF model (KMPN) and a transformer-powered CBF model (NRMS-BERT) were introduced and evaluated.

\item We extend the popular book recommendation dataset Amazon-Book, and have created a brand-new large-scale dataset for movie recommendation.

\item Based on these two realistic datasets, we present extensive experiments showing the effectiveness of our proposed models. We further provide benchmarking results for the newly-collected million-scale dataset which will be shared upon publication.
\end{enumerate}






In addition to providing experimental results showing the power of the models we propose, our aim is to encourage and enable research in content-aware CF recommender systems by providing high-quality datasets and strong models for benchmarking.

\vspace{-0.2cm}
\section{Related Work}
Building upon graph-based recommender systems in which only user-item interactions are exploited~\cite{ying2018graph, he2020lightgcn},
recent advances in KG-based models have demonstrated the effectiveness of fusing valuable external knowledge from auxiliary KGs to improve the both accuracy and explainability of recommendation.
We briefly summarize recent, key methods in KG-based CF models.

\textbf{Embedding-based Methods} rely on the notion that KG head and tail entities are bridged by their first-order relations and that relations with the same type in the KG  share the same embedding.
KG embedding methods (e.g TransE~\cite{bordes2013translating}, TransH~\cite{wang2014knowledge} and TransR~\cite{wang2016text}) are employed to learn KG-aware entity embeddings  which are then fed into CF models for knowledge-aware recommendation~\cite{wang2019multi, zhang2016collaborative, cao2019unifying}.
For example,
KTUP~\cite{cao2019unifying} jointly trained KG completion and item recommendation simultaneously with item embeddings enhanced by TransH-powered KG completion.
MKR~\cite{wang2019multi} employs ``cross\&compress'' units to associate item recommendation in learning KG embeddings.

\textbf{Path-based Methods} focus on longer-range connectivity in the graph. Features are aggregated from distant KG entities to users following the paths that connect them~\cite{hu2018leveraging,jin2020efficient, yu2014personalized, zhao2017meta}.
For example, \citet{yu2014personalized}
and \citet{zhao2017meta} employ manually-designed meta-paths in the heterogeneous KGs to achieve path-aware feature aggregation.
These meta-paths are domain-specific and require domain experts for feature engineering, which is not feasible for larger KGs with their enormous path diversity. 

\textbf{GNN-based Methods} aggregate features grounded in the convolution mechanisms of Graph Neural Networks (GNNs)~\cite{hamilton2017inductive}, including the variants such as Graph Attention Networks (GATs)~\cite{velivckovic2017graph}.
Various systems have been proposed to share neighboring features with nodes of interest~\cite{wang2019knowledge,10.1145/3308558.3313417,jin2020efficient, wang2019kgat, wang2020ckan}.
For example, KGIN~\cite{wang2021learning} embeds relational embeddings in inter-node feature passing to achieve path-aware graph convolution.

In fusing CF and CBF features,
CKE~\cite{zhang2016collaborative} utilize  unsupervised denoising auto-encoders (S-DAE) to obtain textual representations for items.
These features are derived from unsupervised methods, which can be further improved by supervised training and larger language models. In this paper, we introduce NRMS-BERT to obtain more expressive textual representations of items.

In contrast to the above CF models, 
CBF models are extensively investigated in the field of news recommendation since, as noted,  news items expire more quickly than movies and books.
Most research in KG-based CBF focuses on enhancing the item embeddings with KG embeddings by mapping relevant KG entities to the content of items, e.g., by entity linking~\cite{liu2020kred, wang2018dkn}.
For example, DKN~\cite{wang2018dkn} fuses the word-level and knowledge-level representations of news and generates a knowledge-aware embedding vector for content-based recommendation.
However, these methods heavily rely on word-level entity mapping with common knowledge bases, which is prohibited for movies/books since their descriptions mostly consist of imaginary content, such as character names and fictional stories.


\vspace{-0.2cm}
\section{Methodology}
\begin{figure*}
\centering
\includegraphics[width=0.7\linewidth]{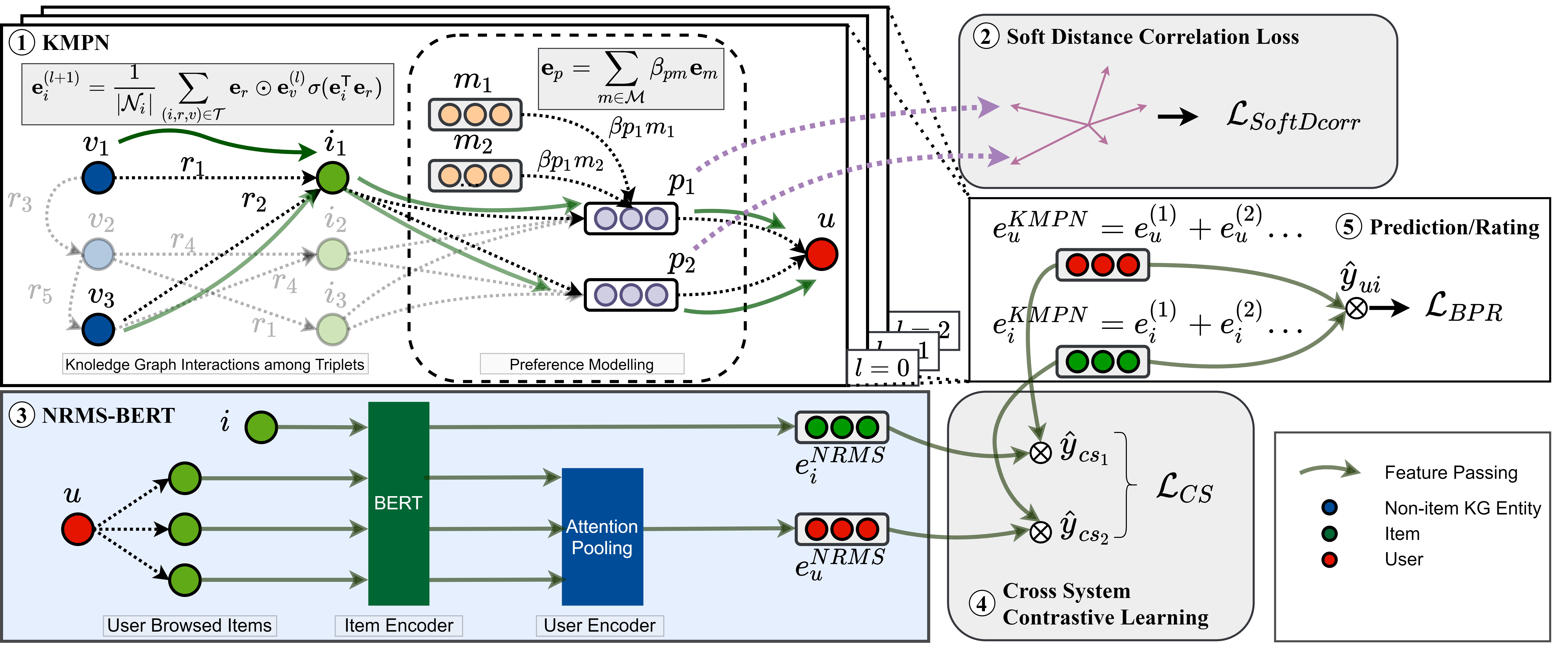}
\vspace{-0.1cm}
\caption{Framework pipeline. (1) KMPN: leverages meta preferences to model users from knowledge graph entities and interacted items; (2) Soft Distance Correlation: encourage preference embeddings to separate at low dimensions; (3) NRMS-BERT: extracts content-based features; (4) Cross System Contrastive Learning: encourages user/item embeddings to learn mutual information from content-based representations;
(5) Rating: uses the dot product of KMPN user/item features.}
\vspace{-0.1cm}
\end{figure*}
In this section, we first introduce notations and data structures (Sec.~\ref{sec:data_structure}).
We describe a CF model that serves as a strong base model (Sec.~\ref{sec:cf_model}), and then proceed to propose a content-based model that rates items through a Transformer with rich texts provided by our datasets (Sec.~\ref{sec:cbf_model}).
Finally, building on top of these two strong models, we introduce a simple but effective fusion framework that leverages the information provided by each component while preserving the CF model as the framework backbone  (Sec.~\ref{sec:merge_model}).

\vspace{-0.2cm}
\subsection{Structured and Unstructured Data}
\label{sec:data_structure}
There are $N_u$ users $\{u| u \in \mathcal{S}_U\}$ and $N_i$ items $\{i|i \in \mathcal{S}_I\}$.
$\mathcal{P}^{+}=\{(u, i) | u\in \mathcal{S}_U, i \in \mathcal{S}_I\}$ is the set of user interactions. Each $(u, i)$ pair indicates that user $u$ interacted with $i$.
Each item $i \in \mathcal{S}_I$ carries unstructured data $\bbx_i$, e.g.  a text description of the item.

The knowledge graph (KG) contains structured information that describes relations between real world entities. 
The KG is represented as a weighted heterogeneous graph 
$\ccalG = (\ccalV, \ccalE)$ with a node set $\ccalV$ consisting of $N_v$ nodes $\{v\}$
and an edge set $\ccalE$ containing all edges between nodes.
The graph is also associated with a relation type mapping function $\phi:\ccalE \rightarrow \ccalR$ that maps each edge to a type in the relation set $\ccalR$ consisting of $N_r$ relations.
Note that all items are included in the KG: $\ccalS_i \subset \ccalV$.

The edges of the knowledge graph are  triplets $\ccalT=\{(h, r, t)|h,t\in \ccalV, r\in \ccalR\}$, where $\ccalV$ is the collection of graph entities/nodes and $\ccalR$ is the relation set.
Each triplet describes that a head entity $h$ is connected to a tail entity $t$ with the relation $r$. For example, $(The\ Shining, film.film.actor, Jack\ Nicholson)$ specifies that Jack Nicholson is a film actor in the movie ``The Shining''.
To fully expose the relationships between heads and tails, the relation set is extended with reversed relation types, i.e., for any $(h, r, t)$ triplet we allow inverse connection $(t, r', h)$ to be built, where $r'$ is the reverse of $r$.  The edge set $\ccalE$ is derived from these triplets.


\subsection{Knowledge-Graph-Enhanced Meta-Preference Network}
\label{sec:cf_model}
We introduce KG-enhanced Meta-Preference Network (thereafter KMPN), a collaborative filtering model.
KMPN aggregates features of all KG entities to items efficiently,  by exploiting relationships in the KG, and then links item features to users for preference-sensitive recommendations.

\vspace{-0.15cm}
\subsubsection{Gated Path Graph Convolution Networks}





Associated with each KG node $v_i$ are feature vectors $\bbe_i^{(0)} \in \bbR^h$.
Each relation type $r \in \ccalR$ is also associated with a relational embedding $\bbe_r$.
A  Gated Path Graph Convolution Networks is a cascade of $L$ convolution layers.
For each KG node, a convolution layer aggregates features from its neighbors as follows:

\vspace{-0.36cm}
\begin{equation}
    \bbe_i^{(l+1)} = \frac{1}{|\ccalN_i|} \sum_{\{v_j|(v_i, r_{ij}, v_j) \in \ccalT\}} \gamma_{ij} \bbe_{r_{ij}} \odot \bbe_j^{(l)},
\end{equation}

where the neighbouring set of $i$: $\ccalN_i=\{v_j|(v_i, r_{ij}, v_j) \in \ccalT\}$, $r_{ij}$ is the type of relation from $v_i$ to $v_j$, and $\gamma_{ij}$ is a gated function that controls messages that flows from $v_j$ to $v_i$:
\vspace{-0.1cm}
\begin{equation}
    \gamma_{ij} = \sigma(\bbe_i ^\mathsf{T} \bbe_{r_{ij}}),
\end{equation}

where $\sigma(.)$ is a sigmoid function that limits the gated value between $0$ and $1$.


As a result, the message passed to a node will be weighted by its importance to the receiving node and the relation type.
Through stacking multiple layers of convolution, the final embedding at a node depends on the path along which the features are shared, as well as the importance of message being transmitted.
To overcome the over-smoothing issue of graph convolutions, the embedding at a KG node after $l$ convolutions is an aggregation of all the intermediate output embeddings:
$
\bbe_i^{l} = \sum_{l'=0}^{l}\bbe_i^{(l')}. 
$

\subsubsection{User Preference modeling}
Inspired by \citet{wang2021learning}, we model users using a combination of preferences.
\citet{wang2021learning} assume that each user is influenced by multiple intents and that each intent is influenced by multiple movie attributes, such as the combination of the two relation types $film.film.director$ and $film.film.genre$.
Based on this assumption, they proposed to aggregate item embeddings to users through ``preferences'', and the embedding of each preference $\bbe_p$ is modelled by all types of edges: $\bbe_p=\sum_{r\in \ccalR} \alpha_{rp} \bbe_r$, where $\alpha_{rp}$ is a Softmax-ed trainable weight and $\bbe_r$ is the embedding of edge relation type $r$.

We take the view that user preferences are not only limited to relations but can be extended to more general cases.
We model each preference $p$ through a combination of a set of meta-preferences $\ccalM$ with in total $N_m$ meta-preferences:
each meta-preference $m \in \ccalM$ is associated with a trainable embedding $\bbe_m \in \bbR^{h}$, and a preference $p$ is formed by these meta-preferences as follows:
\vspace{-0.1cm}
\begin{equation}
    \bbe_p=\sum_{m\in \ccalM} \beta_{pm} \bbe_m,
    \vspace{-0.1cm}
\end{equation}

where the linear weights $\{\beta_{pm}|m \in \ccalM\}$ is derived from trainable weights $\{\hat{\beta}_{pm}|m \in \ccalM\}$ for each preference $p$:
\vspace{-0.1cm}
\begin{equation}
    \beta_{pm} = \frac{\exp{(\hat{\beta}_{pm})}}{\sum_{m'\in \ccalM}\exp{(\hat{\beta}_{pm'})}}.
    \vspace{-0.1cm}
\end{equation}

As a result, meta-preferences reflect the general interests of all users.
A particular user can be profiled by aggregating the embeddings of interacted items through these preferences:
\vspace{-0.1cm}
\begin{equation}
    \bbe_u^{(l)} = \sum_{p\in \ccalP} \alpha_p \sum_{(u,i) \in \ccalP^{+}} \bbe_i^{(l)} \odot \bbe_p,
    \vspace{-0.1cm}
\end{equation}
where $\ccalP$ is the collection of $N_p$ preferences $\{p\}$ and $\alpha_p$ is an attention mechanism that weights the interest of users over different preferences:
\vspace{-0.3cm}
\begin{equation}
    \alpha_p = \frac{\exp{(\bbe_p ^\mathsf{T} \bbe_u)}}{\sum_{p'\in \ccalP}\exp{(\bbe_{p'} ^\mathsf{T} \bbe_u)}}.
    \vspace{-0.1cm}
\end{equation}

In summary, each preference is formed by general and diverse meta-preferences, and users are further profiled by multiple preferences that focus on different aspects of item features.
As with items, the final user embedding is:
$
\bbe_u^{l} = \sum_{l'=0}^{l}\bbe_u^{(l')}.
$
\subsubsection{Soft Distance Correlation for Training Preferences}
Having modelled users through preferences, \citet{wang2021learning} added an additional loss that utilizes Distance Correlation (DCorr) to separate the representations of these learnt preferences as much as possible, in order to obtain diverse proxies bridging users and items.
Though the authors demonstrate a considerable improvement over baselines, we take the view that applying constraints to all dimensions of preference embeddings restricts their expressiveness, as they are trained to be very dissimilar and have diverse orientations in latent space.

We adopt a softer approach: \textbf{Soft Distance Correlation Loss}, which firstly lowers the dimensionality of preference embeddings with \textit{Principal Component Analysis}  (PCA)~\cite{hotelling1933analysis} while keeping the most differentiable features in embeddings, and then applies distance correlation constraints to encourage diverse expression in lower dimensions:
\vspace{-0.1cm}
\begin{equation}
    \hat{\bbe}_p = \mathcal{PCA}(\{\bbe_{p'}|p'\in \ccalP \}) \in \bbR^{h\epsilon};
    \vspace{-0.1cm}
\end{equation}
\vspace{-0.1cm}
\begin{equation}
    \ccalL_{SoftDCorr} = \sum_{p,p'\in \ccalP, p\neq p'} \frac{DCov(\hat{\bbe}_p, \hat{\bbe}_{p'})}{\sqrt{DVar(\hat{\bbe}_p) \cdot  DVar(\hat{\bbe}_{p'})}}.
    \vspace{-0.1cm}
\end{equation}

where $\epsilon$ is the ratio of principal components to keep after PCA,
and $DCov(\cdot)$ computes distance covariance while $DVar(\cdot)$ measures distance variance.

Through encouraging diverse expression at only lower dimensions, preferences have retained the flexibility in higher dimensions.
Of course, $\epsilon=1$ yields the original Distance Correlation Loss.

\subsubsection{Model Optimization with Reciprocal Ratio Negative Sampling}

Following the common approach, the dot product between user and item embeddings is used for rating:
$
\hat{y}_{ui} = (\bbe_u^{L})^\mathsf{T} \cdot \bbe_i^{L}.
$

Both of the datasets we study do not provide hard negative samples: i.e., we do not have samples of items with which viewers chose not to interact.
A common practice to synthesize negative examples is to randomly sample from users' unobserved counterparts $\ccalP^{-}=\{(u, i^{-})|(u, i^{-}) \notin \ccalP^{+}\}$.
However,  an item is not necessarily ``not interesting'' to a user if no interaction happens, as not all items have been viewed.
We, therefore, propose to adopt \textit{Reciprocal Ratio Negative Sampling}, where items with more user interactions are considered popular and are sampled less frequently based on the assumption that popular items are less likely to be hard negative samples for any user.
The sampling distribution is given by a normalized reciprocal ratio of item interactions:
\vspace{-0.1cm}
\begin{equation}
    i^{-} \sim P(i) \propto \frac{1}{c(i)} \mbox{ for } i \in S_I
    \vspace{-0.1cm}
\end{equation}

where $c(i)$ counts the interactions of all users with the item $i$.

The training set therefore consists of positive and negative samples: $\ccalU = \{(u, i^{+}, i^{-})|(u, i^{+}) \in \ccalP^{+}, (u, i^{-}) \in \ccalP^{-}\}$.
Pairwise BPR loss~\cite{rendle2012bpr} is adopted to train the model, which exploits a contrastive learning concept to assign higher scores to users' browsed items than those items that users are not interested in:
\vspace{-0.1cm}
\begin{equation}
    \ccalL_{BPR}=\sum_{(u, i^{+}, i^{-})\in \ccalU} -\ln (\sigma(\hat{y}_{ui^{+}}-\hat{y}_{ui^{-}})).
    \vspace{-0.1cm}
\end{equation}

Together with commonly-used embedding L2 regularization and Soft Distance Correlation loss, the final loss is given by:
\vspace{-0.1cm}
\begin{equation}
    \ccalL_{KMPN} = \ccalL_{BPR} + \lambda_{1} \frac{1}{2}||\Theta||_2^2 + \lambda_{2}\ccalL_{SoftDCorr},
    \vspace{-0.1cm}
\label{eqn:cf_loss}
\end{equation}
where $\Theta=\{\bbe_u^L, \bbe_{i^{+}}^L, \bbe_{i^{-}}^L|(u, i^{+}, i^{-})\in \ccalU \}$, and $||\Theta||_2^2$ is the L2-norm of user/item embeddings. $\lambda_1$ and $\lambda_2$ are hyperparameters that control loss weights.

\subsection{Neural Recommendation with Multi-Head Self-Attention enhanced by BERT}
\label{sec:cbf_model}
Inspired by NRMS \cite{wu-etal-2019-neural-news}, a powerful content-based system that exhibited outstanding performance in news recommendation,
we propose a variant of NRMS, \textbf{NRMS-BERT}, that further utilizes a pre-trained transformer for extracting contextual information from descriptions of items.

The model consists of an item encoder and a user encoder.
Rather than train self-attention blocks from scratch as in NRMS, NRMS-BERT incorporates a pre-trained BERT as the base block for language understanding.

\subsubsection{Item Encoder}
The item encoder encodes the text description string $\bbx_i$ of any item $i\in \ccalS_i$ through BERT into embeddings of size $h$ by extracting the embeddings of \texttt{<CLS>} at the last layer:
\vspace{-0.1cm}
\begin{equation}
    \bbe_i = \mathcal{BERT}(\bbx_i) \in \bbR^h.
    \vspace{-0.1cm}
\end{equation}

For each user, the item encoder encodes one positive item $\bbe_{i^{+}}$ and $K$ negative items $\bbe_{i^{-}_1}$, ..., $\bbe_{i^{-}_{K}}$.
$B$ items are randomly sampled from the user's browsed items $i_{u,1}$,...,$i_{u,B}$.
These browsed items are encoded and gathered to $\bbE_u = [\bbe_{i_{u,1}},...,\bbe_{i_{u,B}}] \in \bbR^{B \times h}$.
\subsubsection{User Encoder}
The user encoder uses items with which users interacted to produce a content-aware user representation.
The final user representation is a weighted sum of the $B$ browsed items:
\vspace{-0.2cm}
\begin{equation}
    \bbe_u = \sum_{b=1}^{B} \alpha_{b} \bbe_{i_{u,b}}
    \vspace{-0cm}
\end{equation}

where $\alpha_b$ is the attention weight assigned to $i_{u,b}$ obtained by passing features through two linear layers:
\vspace{-0.1cm}
\begin{equation}
    \alpha_b = \frac{\exp{(\hat{\bbA}_b)}}{\sum_{b'=1,..,B}\exp{(\hat{\bbA}_{b'})}};
    \vspace{-0.1cm}
\end{equation}

\vspace{-0.1cm}
\begin{equation}
    \hat{\bbA} = \tanh(\bbE_u \bbA_{fc_1}+\bbb_{fc_1})\bbA_{fc_2}+\bbb_{fc_2} \in \bbR^{B \times 1}
\end{equation}

where $\bbA_{fc_1} \in \bbR^{h\times \frac{1}{2}h}$, $\bbb_{fc_1} \in \bbR^{\frac{1}{2}h}$, $\bbA_{fc_2} \in \bbR^{\frac{1}{2}h\times 1}$, and $\bbb_{fc_2} \in \bbR^{1}$ are weights and biases of two fully-connected layers, respectively.

\subsubsection{Model Optimization}
The rating is the dot product of user and item embeddings:
$
\hat{y}_{ui} = (\bbe_u)^\mathsf{T} \cdot \bbe_i.
$
Assume that the scores of the positive samples and negative samples are $\hat{y}^{+}$ and $\hat{y}^{-}_{1}$,...,$\hat{y}^{-}_{K}$, following \cite{wu-etal-2019-neural-news}, the loss for optimization is the log posterior click probability of item $i$:
\vspace{-0.1cm}
\begin{equation}
    \ccalL_{NRMS} = -\sum_{i \in \ccalS_i} \log{(\frac{\exp{(\hat{y}^{+})}}{\exp{(\hat{y}^{+})}+\sum_{k=1,..,K}\exp{(\hat{y}^{-}_{k})}})}
    \vspace{-0.1cm}
\end{equation}

\subsection{Fusion of CF and CBF: Content-Aware KMPN}
\label{sec:merge_model}
To fuse the information from a CBF model (NRMS-BERT) to a CF model (KMPN), we must bridge some inconsistencies between the two types of models.
CBF models that utilize large transformers cannot be co-optimized with KG-based CF models, as graph convolution requires all embeddings to be present before convolution and this requires enormous GPU memory for even one single forward pass.
As a result, a more efficient solution merges the pre-trained CBF features into the training of KG-CF component, enriching the learned representations.

Conventional fusing approaches, including (1) concatenating CBF features with CF before or after graph operations (early fusion and late fusion) and (2) using an auxiliary loss to reduce the distance between CBF and CF user/item features, are appealing and thus serve as our baselines in Sec.~\ref{sec:experiments:cross_system_learning}.



In line with our aim to use a CF model for movie and book recommendations, we present a novel and efficient approach for training a better KMPN: \textit{Cross-System Contrastive Learning}.
KMPN is still used as the backbone and it is trained with the aid of a pre-trained NRMS-BERT.

In KMPN training,
for users and items in $(u, i^{+}, i^{-})\in \ccalU$,
embeddings are generated from NRMS-BERT: $e_u^{NRMS}$, $e_{i^{+}}^{NRMS}$, $e_{i^{-}}^{NRMS}$, and from KMPN: $e_u^{KMPN}$, $e_{i^{+}}^{KMPN}$, and $e_{i^{-}}^{KMPN}$.

\textit{Cross-System Contrastive Loss} is adopted to encourage KMPN system to learn to incorporate content-sensitive features from NRMS-BERT features:
\begin{equation}
\begin{aligned}
    \ccalL_{CS}=\sum_{(u, i^{+}, i^{-})\in \ccalU} &-\ln \Big (\sigma \big ((\bbe_u^{KMPN})^\mathsf{T} \cdot (\bbe_{i^{+}}^{NRMS}-\bbe_{i^{-}}^{NRMS})\big) \Big) \\
    &-\ln \Big (\sigma \big ((\bbe_u^{NRMS})^\mathsf{T} \cdot (\bbe_{i^{+}}^{KMPN}- \bbe_{i^{-}}^{KMPN})\big) \Big)
\end{aligned}
\label{eqn:cross_system_loss}
\end{equation}

This loss encourages KMPN to produce item embeddings that interact not only with KMPN's own user embeddings, but also with NRMS-BERT's user embeddings.
Similarly, user embeddings of KMPN are trained to interact with items of NRMS-BERT.
This allows $\bbe_i^{KMPN}$ to learn mutual expressiveness with $\bbe_i^{NRMS}$, but without approaching the two embeddings directly using similarity (e.g. cosine-similarity), which we found not to work well (discussed in Sec.~\ref{sec:experiments:cross_system_learning}).
In this case, $\bbe_u^{NRMS}$ serves as an `anchor' with which the item embeddings of two systems learn to share commons and increase their mutuality.
This loss encourages $\bbe_i^{KMPN}$ and $\bbe_i^{NRMS}$ to lie in the same hidden space hyperplane on which features have the same dot-product results with $\bbe_u^{NRMS}$.
This constraint encourages KMPN to grow embeddings in the same region of hidden space, leading to mutual expressiveness across the two systems.
Finally, the optimization target is:
\vspace{-0.1cm}
\begin{equation}
    \ccalL_{CKMPN} = \ccalL_{KMPN} + \lambda_{CS}\ccalL_{CS},
\vspace{-0.1cm}
\end{equation}

where $\lambda_{CS}$ controls the weight of the Cross-System Contrastive Loss.

\vspace{-0.1cm}
\subsection{Training Details}
All experiments were run on 8 $\times$ NVIDIA A100 Tensor Core GPUs with batch size $8192\times 8$ for KMPN/CKMPN and $4\times8$ for NRMS-BERT.
KMPN/CKMPN's optimizer was Adam \cite{DBLP:journals/corr/KingmaB14} with 2000 epochs of linearly decayed learning rate from $10^{-3}$ to $0$ for Amazon-Book-Extended and $5\times10^{-4}$ to $0$ for Movie-KG-Dataset.
NRMS-BERT was optimized with Adam at a constant learning rate of $10^{-4}$.


\section{Datasets}
For the purpose of evaluating the content-aware fusion network proposed in this paper, we introduce two new datasets for recommendation: 
(1) a large-scale high-quality movie recommendation dataset that was newly collected from real user behaviors (Sec.~\ref{sec:movie_dataset});
(2) the Amazon-Book-Extended dataset based on the Amazon-Book dataset~\cite{wang2019kgat} and newly augmented with textual book descriptions to allow the evaluation of content-based systems (Sec.~\ref{sec:amazon_dataset}).

\vspace{-0.2cm}
\subsection{Movie KG Dataset}
\label{sec:movie_dataset}
\subsubsection{Dataset Construction}
This dataset was formed by real user browsing behaviors that were logged by a popular commercial browser between 13/05/2021 and 20/06/2021 (39 days in total).
Sensitivity of data and privacy was firstly removed by decoupling users' real identity (e.g. IP, account identifier) from the data and assigning each user an insensitive unique virtual identifier.
To link users' browsing history with movies, we exploited a commercial knowledge graph that consists of a large number of entities intended to cover the domain and a rich set of relationships among them.
Entities with type ``movie'' were extracted together with their one-hop neighbors and edges.
In this way a movie-specific sub-graph was extracted from the original knowledge graph.

Drawing from billions of user browsing logs, knowledge-graph entity linking was performed to match web page titles with the movie titles in the movie-specific KG. 
Movie entities were extracted using an internal service that provides Named Entity Recognition and entity linking.
In order to make the dataset more compact:
(1) consecutive and duplicated user interactions with the same movie were merged to one single record and their browsing times (DwellTime) was aggregated;
(2) less active users with fewer than 10 interactions in 39 days were dropped;
(3) the most frequent 50,000 movies with the most user interactions were selected, since records of tail movies with few interactions were considered unreliable.

125,218 active users and 50,000 popular movies were chosen.
KG entities related to these movies were extracted to form a sub-KG with 25,0327 entities (including movie items).
Each user is represented by [$UserID$, $UserHistory$], where $UserID$ is a unique identifier that has been delinked from real user identity, and $UserHistory$ is an ID list of movie items that the user has browsed.

For training and evaluating recommender systems, $UserHistory$ was split into train/validation/test sets:
the first 80\% of users' historical interactions (ordered by click date-time) are in the train set; 80\% to 90\% serves as the validation set; and the remaining is reserved for testing.
During validation and test, interactions in the train set serve as users' previous clicked items, and systems are evaluated on their ratings for test set items.
In addition to the traditional data split strategy, a \textbf{cold-start user set} was created to evaluate model performance on users outside the training dataset.
3\% of users were moved to the cold-start set and they are not available in model training.
This portion of interactions was split into a cold-start history set with first 80\% interactions of each user, and a test set with the remaining 20\%.

A comparison with some popular existing recommendation datasets is provided in Table~\ref{tab:comparison_with_datasets}.
\textbf{MovieLens-20M}\footnote{\href{https://grouplens.org/datasets/movielens/}{https://grouplens.org/datasets/movielens/}} is a popular benchmark that has been widely used, while
\textbf{Flixster}\footnote{\href{https://sites.google.com/view/mohsenjamali/flixter-data-set}{https://sites.google.com/view/mohsenjamali/flixter-data-set}} is large but less popular in recent work.
\textbf{Amazon-Book}~\cite{wang2019kgat}, \textbf{Last-FM}~\cite{wang2019kgat}, \textbf{Book-Crossing}~\cite{ziegler2005improving}\footnote{\href{http://www2.informatik.uni-freiburg.de/~cziegler/BX/}{http://www2.informatik.uni-freiburg.de/~cziegler/BX/}} and \textbf{Alibaba-iFashion}~\cite{wang2021learning} are dedicated for KG-based CF model evaluation.

In summary,  our new movie dataset is based on large user populations and movie inventory and draws from an extensive movie-specific knowledge graph with millions of triplets and entities.   This is a rich resource not yet provided by existing popular movie recommendation datasets.
Our dataset provides not only titles and genres (as in MovieLens-20M), but also other content-based information, such as rich summary texts of movies that enable content-based recommendation with large language understanding models such as BERT and GPT-2.

\begin{table*}[!ht]
\caption{Comparison with existing popular datasets for recommender systems.}
\label{tab:comparison_with_datasets}
\vspace{-0.2cm}
\small
\begin{tabular}{llllllllll}
\toprule
Type     & Dataset                 & \#Users & \#Items & \#Interactions & KG & \#Entities & \#Relations & \#Triplets &  Descriptions           \\ 
\midrule
Movie    & MovieLens-20M           & 138,000 & 27,000  & 20,000,000     & No              & N/A        & N/A         &    N/A        & No                      \\
Moive   & Flixster  & 1,002,796   & 66,730    & 1,048,576   & No    & N/A &   N/A & N/A & No\\
Book     & Amazon-Book             & 70,679  & 24,915  & 847,733        & Yes             & 88,572     & 39          & 2,557,746  & No                      \\ 
Book    & Book-Crossing   & 276,271   & 271,379  & 1,048,575 & Yes & 25,787 & 18 & 60,787 & No\\
Music    & Last-FM                 & 23,566  & 48,123  & 3,034,796      & Yes             & 58,266     & 9           & 464,567    & No                      \\ 
Shopping & Alibaba-iFashion        & 114,737 & 30,040  & 1,781,093      & Yes             & 59,156     & 51          & 279,155    & No                      \\
\midrule
Book     & Amazon-Book-Extended    & 70,679  & 24,915  & 847,733        & Yes             & 88,572     & 39          & 2,557,746  &  Yes       \\ 
Movie    & Movie-KG-Dataset & 125,218 & 50,000  &  4,095,146              & Yes             & 250,327    & 12          & 12,055,581 & Yes\\
\bottomrule
\end{tabular}
\vspace{-0.3cm}
\end{table*}

\subsubsection{Statistical Analysis}
\label{sec:movie_statistical_analysis}
\begin{figure}
\centering     
\subfigure[Length of user history ($\mu=32.70$, $\sigma=45.76$)]{\label{fig:hist_number_user_interactions}\includegraphics[width=0.45\linewidth]{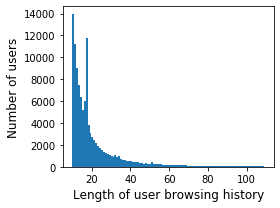}}
\hfill
\subfigure[Users' interactions with each movie ($\mu=83.33$, $\sigma=582.50$)]{\label{fig:hist_movie_interaction_count}\includegraphics[width=0.45\linewidth]{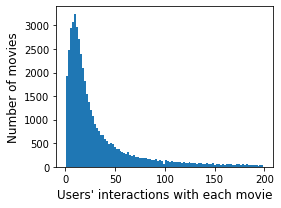}}\\
\subfigure[Length of movie description in Movie-KG-Dataset ($\mu=425.22$, $\sigma=329.93$)]{\label{fig:hist_movie_text_length}\includegraphics[width=0.45\linewidth]{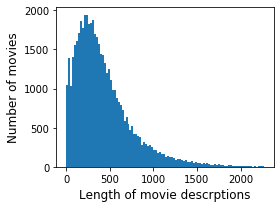}}
\hfill
\subfigure[Length of book description in Amazon-Book-Extended ($\mu=151.75$, $\sigma=206.98$)]{\label{fig:hist_book_text_length}\includegraphics[width=0.45\linewidth]{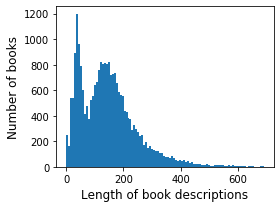}}
\vspace{-0.4cm}
\caption{Key statistics of Movie-KG-Dataset (Fig.~\ref{fig:hist_number_user_interactions}-\ref{fig:hist_movie_text_length}) and  Amazon-Book-Extended (Fig.~\ref{fig:hist_book_text_length}).}
\label{fig:dataset_analysis}
\vspace{-0.4cm}
\end{figure}
Fig.~\ref{tab:comparison_with_datasets} presents the key statistics of the dataset.
This dataset contains 125,218 users, 50,000 movie items, and 4M+ interactions. The knowledge graph contains 250,327 entities with 12 relation types and 12,055,581 triplets.
The attributes of movies include content-based properties (title, description, movie length) and properties already tied to knowledge graph entities (production company, country, language, producer, director, genre, rating, editor, writer, honors, actor).
Fig.~\ref{fig:hist_number_user_interactions}-\ref{fig:hist_movie_text_length} shows the distribution of length of interactions with movies per user, number of interactions with users per movie, and the length of movie descriptions.
The average length of movie descriptions is 425 words, which is long enough for models to encode movies from the summary of stories.
The average number of user interactions is 32.70, providing rich resources for modeling long-term user interest.
The first 19 days of data (with similar data distribution but fewer users and interactions) is separated as ``Standard'' for most research purposes, while the full 39 days' are released as ``Extended'' for more extended use, such as training industrial products.

\vspace{-0.2cm}
\subsection{Extended Amazon Book Dataset}
\label{sec:amazon_dataset}
The Amazon-Book dataset was originally released by~\cite{wang2019kgat} and has been used in the development of many advanced recommendation systems.
However, this dataset provides only CF interaction data without content-based features needed for CBF evaluation.
To fill this gap, we extend the dataset with descriptions extracted from multiple data sources, noting that the dataset was originally collected in 2014 and many items have since expired on the Amazon website.   The procedure is summarized as follows:

(1) We matched product descriptions in Amazon with entries in Amazon-Book using their unique item identifiers (asin). 24,555 items were successfully matched (98.56\%);
(2) we then matched each of the remaining items with the most relevant entry in a huge commercial KG which enables rich descriptions to be extracted. 325 items were matched at this step;
(3) the remaining 28 items were matched manually to the most relevant products in Amazon.

The distribution of description lengths is given in Fig.~\ref{fig:hist_book_text_length}, with a mean length of 151.75 words.

\section{Experiments}

\begin{table*}[!ht]
\caption{Comparison of baseline models and our proposed models on Amazon-Book-Extended dataset. Numbers \underline{underlined} are existing state-of-the-art performance, while best performance of the proposed models is marked in \textbf{bold}.
Relative improvements over baseline models are shown in the last row.
The average of 3 runs is reported to mitigate experimental randomness.
All metrics of KMPN are significantly higher than KGIN ($p<0.05$).
Metrics with (*) are significantly higher than KMPN ($p<0.05$).
}
\label{tab:compare_baselines}
\vspace{-0.3cm}
\small
\centering
\begin{tabular}{llllllllll}
\toprule
                                  & \multicolumn{3}{c}{Recall} & \multicolumn{3}{c}{ndcg}   & \multicolumn{3}{c}{Hit Ratio}  \\
~                                 & @20   & @60   & @100    & @20   & @60     & @100  & @20   & @60   & @100        \\
\midrule
BPRMF                                & 0.1352&	0.2433&	0.3088&	0.0696&	0.09568&	0.1089&	0.2376&	0.3984&	0.4816

          \\
CKE                               & 0.1347&	0.2413&	0.3070&	0.0691&	0.09482&	0.1081&	0.2373&	0.3963&	0.4800
            \\
KGAT                              & 0.1527&	0.2595&	0.3227&	0.0807&	0.10655&	0.1194&	0.2602&	0.4156&	0.4931
       \\
KGIN                              & \underline{0.1654}&	\underline{0.2691}&	\underline{0.3298}&	\underline{0.0893}&	\underline{0.1145}&	\underline{0.1267}&	\underline{0.2805}&	\underline{0.4289}&	\underline{0.5040}
       \\
\midrule
KMPN & 0.1719 & 0.2793 & 0.3405   & 0.0931 & 0.1189 & 0.1315 & 0.2910 & 0.4421 & 0.5166       \\
w/o Soft Distance Correlation (DCorr)  & 0.1704 & 0.2790 & 0.3396   & 0.0924 & 0.1185  & 0.1310 & 0.2881 & 0.4419 & 0.5152       \\
w/o Soft DCorr and Reciprocal Ratio Negative Sampling                              & 0.1690 & 0.2774 & 0.3391   & 0.0913 & 0.1177  & 0.1302 & 0.2872 & 0.4414 & 0.5155       \\
\midrule
NRMS-BERT &	0.1142 &	0.2083	& 0.2671 &	0.0592 &	0.0817 &	0.0935 &	0.2057 &	0.3487 &	0.4273 \\
\midrule
Mixture of Expert (KMPN + NRMS-BERT)& \textbf{0.1723}&	0.2791&	0.3281&	\textbf{0.0933}&	0.1161&	0.1214&	\textbf{0.2913}&	0.4425&	0.5022\\
CKMPN ($\lambda_{CS}=0.2$) & 0.1699 & 0.2812 & \textbf{0.3461}   & 0.0922 & 0.1190 & 0.1319 & 0.2880 & 0.4460 & 0.5235       \\
CKMPN ($\lambda_{CS}=0.1$) & 0.1718 & \textbf{0.2821}* & 0.3460*   & 0.0928 & \textbf{0.1197}* & \textbf{0.1326}* & 0.2908 & \textbf{0.4474}* & \textbf{0.5244}*   \\
\midrule
Improv. (\%) CKMPN v.s. Baselines &	3.90&	4.82&	4.94&	4.31&	4.55&	4.59&	3.72&	4.33&	4.04
\\
\bottomrule
\end{tabular}
\end{table*}
In this section, we present experimental results of the three proposed models: KMPN, NRMS-BERT, and CKMPN, with detailed discussions.
We firstly demonstrate the improvement brought by each component over existing baseline models including state-of-the-art models on Amazon-Book-Extended, which has an extensive record of publications that enables direct and fair comparison.
We also provide benchmark results on our brand-new Movie-KG-Dataset to enable the development and comparison of future models.

\vspace{-0.1cm}
\subsection{Evaluation Metrics}

Following common practice~\cite{he2017neural, wang2019kgat, wang2021learning}, we report metrics (summarized in \cite{krichene2020sampled}) for evaluating model performance:
(1) \textit{Recall@K}: within top-$K$ recommendations, how well the system recalls the test-set browsed items for each user;
(2) \textit{ndcg@K} (Normalized Discounted Cumulative Gain): increases when relevant items appear earlier in the recommended list; 
(3) \textit{HitRatio@K}: how likely a  user finds at least one interesting item  in the recommended top-K items.


\vspace{-0.15cm}
\subsection{Baselines}
We take the performance of several recently published recommender systems as points for comparison.
We carefully reproduced all these baseline systems from their repositories\footnote{As a result, the results reported here may differ from those of the original papers.}.

\textbf{BPRMF}~\cite{rendle2012bpr}: a powerful matrix factorization method that applies a generic optimization criterion BPR-Opt for personalized ranking.
Limited by space, other matrix factorization based methods are not presented since BPRMF outperformed them.

\textbf{CKE}~\cite{zhang2016collaborative}: a CF model that leverages heterogeneous information in a knowledge base for recommendation.

\textbf{KGAT}~\cite{wang2019kgat}: Knowledge Graph Attention Network (KGAT) which explicitly models high-order KG connectivities in KG. The models' user/item embeddings were initialized from the pre-trained \textbf{BPRMF} weights.

\textbf{KGIN}~\cite{wang2021learning}: a state-of-the-art CF model that models users' latent intent as a combination of KG relations.

\textbf{Mixture of Expert}: the output scores of two systems, KMPN and NRMS-BERT, passed through 3 layers of a Multi-Layer Perception (MLP) to obtain final item ratings.



\vspace{-0.2cm}
\subsection{Model Performance on Amazon Dataset}

\textbf{Comparison with Baselines:}
Performance of models is presented in Table~\ref{tab:compare_baselines}. 
Our proposed CF model, KMPN, with \textit{Reciprocal Ratio Negative Sampling} and \textit{Soft Distance Contrastive Learning}, achieved a substantial improvement on all metrics over the performance of the existing state-of-the-art model KGIN; for example, Recall@20 was improved from 0.1654 to 0.1719, Recall@100 from 0.3298 to 0.3405, and ndcg@100 from 0.1267 to 0.1315, with statistical significance $p<0.05$.
\textbf{The statistical significance of all relative improvements mentioned in this paper is verified ($p<0.05$).}


NRMS-BERT models user-item preferences using only item summary texts, without external information from a knowledge base.
It still achieves 0.1142 in Recall@20 and 0.4273 Hit Ratio@100, not far from the KGIN baseline at 0.5040 Hit Ratio@100. 
This suggests that our extension to Amazon-Book dataset serves as a good additional content-based source for constructing recommendation systems.

\begin{table}[!ht]
\vspace{-0.1cm}
\caption{Results of conventional feature fusion approaches on Amazon-Book-Extended dataset.
}
\label{tab:comparing_conventional_fusion}
\vspace{-0.3cm}
\centering
\small
\begin{tabular}{lccc}
\toprule
Fusion Approach & \multicolumn{1}{c}{Recall@60} & \multicolumn{1}{c}{ndcg@60}   & \multicolumn{1}{c}{Hit Ratio@60}  \\

\midrule
Early Fusion (concat) & 0.2708 &	0.1148 &	0.4299
\\
Late Fusion (concat+linear)  & 0.2769 &	0.1164 &	0.4381

\\
Late Fusion (MultiHeadAtt)  & 0.2778 &	0.1175 &	0.4385
\\
Cos-Sim & 0.2436 & 0.1026
 & 0.4001 \\
CKMPN (ours) & \textbf{0.2821} & \textbf{0.1197} & \textbf{0.4474}\\
\bottomrule
\end{tabular}
\vspace{-0.1cm}
\end{table}

\textbf{Comparison with Conventional Fusion Approaches:}
Conventional feature fusion methods are convenient options for fusing the features of one system into the training of another.
We compare these conventional approaches with our proposed \textit{Cross-System Contrastive Learning} (Cross-System CL) in Table~\ref{tab:comparing_conventional_fusion}:

\begin{itemize}
    \item \textbf{Early Fusion}: CBF features are concatenated to the trainable user/item embeddings of KMPN before the graph convolution layers.
    \item \textbf{Late Fusion}: CBF features are fused to the output user/item embeddings of KMPN after the graph convolution layers.
    Many feature aggregation methods were experimented and the best of them are reported in Table~\ref{tab:comparing_conventional_fusion}: (1) concat+linear: CF features are concatenated with CBF features, and they pass through 3 layers of Multi-layer Perceptron into embeddings of size $\mathbf{R}^{2\times h}$. 
    (2) MultiHeadAtt: CF and CBF features passed through 3 Multi-head Self-Attention blocks into embeddings of size $\mathbf{R}^{2\times h}$.
    \item \textbf{Cos-Sim}: An auxiliary loss grounded on cosine-similarity is incorporated in training to encourage the user/item embeddings of KMPN to approach those of NRMS-BERT.
\end{itemize}

It can be concluded that these Na\"ive feature aggregation approaches do not perform well in fusing pre-trained CBF features into CF training.
(1) The performance of \textbf{Late Fusion} shows that when the already-learned NRMS-BERT item/user embeddings pass through new layers, these layers undid the learned representations from NRMS-BERT and led to  only degraded performance.
(2) \textbf{Cos-Sim} shows that the auxiliary loss based on cosine-similarity places a reliance on NRMS-BERT's features, which damages the KMPN training by limiting the expressiveness of KMPN to that of NRMS-BERT. As a result, the performance is decreased from 0.2793 (KMPN) to 0.2436 (Cos-Sim) recall@60. 


In contrast, our proposed Cross-System CL approach substantially improves all @60/@100 metrics while keeping the model's performance of @20.
For example, as shown, CKMPN achieved 0.1718 in Recall@20, the same as that of KMPN.
But CKMPN (0.3461 Recall@100) outperforms KMPN (0.3405 Recall@100) by 1.6\% with statistical significance $p<0.05$.
This demonstrates that gathering item and user embeddings from one system (KMPN) with those of the other system (NRMS-BERT), through proxies (Cross-System CL), encourages KMPN to learn and fuse content-aware information from the learned representations of a CBF model.
Cross-System CL complements the aforementioned shortages of conventional feature fusing methods by merging features without breaking the already-learned representation and without directly approaching two systems' outputs.

Finally, compared to the Mixture of Expert model (Table~\ref{tab:compare_baselines}), where scores of two systems are directly merged by MLP layers, CKMPN achieved generally better performance in @60/100 results, showing that our method is an in-depth collaboration of two systems instead of a simple aggregation of system outputs.
In conclusion, the fusion of these content-based features significantly enhanced KMPN's ability to present more relevant items in the top-$100$ recommendation list.
A case study is presented later in Sec.~\ref{sec:movie_experiments}.

In the following subsections, to support the rationale of our designs, we provide detailed discussions exploring the effects brought by each proposed component.

\subsubsection{Effects of Reciprocal Ratio Negative Sampling}
\label{sec:experiments:negative_sampling}
Through \textit{Reciprocal Ratio Negative Sampling}, KMPN further improves the metrics: Recall@20 increases from 0.1690 to 0.1704 as shown in Table~\ref{tab:compare_baselines}.
In line with our intuition, reducing the probability of sampling popular items as negative samples for training can yield benefits in model learning.
This demonstrates that while viewed-but-not-clicked (hard negative) samples are not available to the model, \textit{Reciprocal Ratio Negative Sampling} enhances the quality of negative samples.

\subsubsection{Effects of Meta Preferences}
\label{sec:experiments:meta_preferences}

\begin{figure}
\centering     

\subfigure[Model performance varies with the number of meta-preferences. Best performance is obtained at $N_m=64$.]{\label{fig:effects_of_meta_preference}\includegraphics[width=0.49\linewidth]{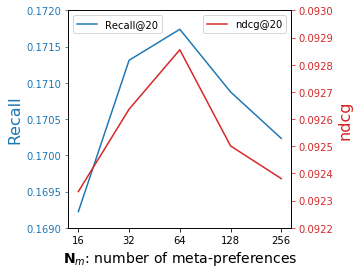}}
\hfill
\subfigure[Model performance varies with the ratio $\epsilon$ of Soft Distance Correlation (DCorr).
$\epsilon=0$ trains preference embeddings without DCorr Loss; $\epsilon=1$ reduces to standard DCorr loss.
]{\label{fig:effects_of_soft_distance_correlation}\includegraphics[width=0.49\linewidth]{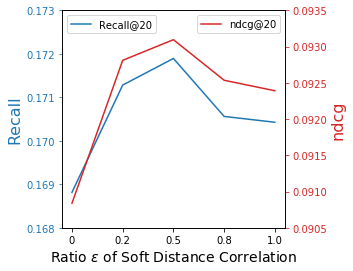}}
\vspace{-0.5cm}
\caption{Hyperparameter evaluation with number of meta-preferences and Soft Distance Correlation.}
\label{fig:effects_hyperparameters}
\vspace{-0.6cm}
\end{figure}

An obvious research question is how the design of modeling users through meta-preferences improves the model performance.

As shown in Fig.~\ref{fig:effects_of_meta_preference}, the system performance against different $N_m$ (number of meta preferences) is plotted.
$N_m=16$ achieved worse performance than $N_m\geq32$ since a small number of meta preferences limited the model's capacity of modeling users.
The performance on all metrics increases until it peaks at $N_m=64$, and then it starts to decrease at $N_m\geq128$.
This suggests that bringing too many meta preferences induces overfitting and will not further improve the system.
It is a good model property in practice since a moderate $N_m=64$ is sufficient for achieving the best performance.

\subsubsection{Effects of Soft Distance Correlation Loss}
\label{sec:experiments:soft_distance_correlation}
The hyperparameter $\epsilon$ controls the number of principal components to keep after PCA dimension reduction.
The lower the ratio, the more flexibility the preference embeddings will recover in dimensions from the standard Distance Correlation constraint.
$\epsilon=0$ removes the Soft Distance Correlation Constraint completely, while $\epsilon=1$ reduces to a standard Distance Correlation Loss. 
As shown in Fig.~\ref{fig:effects_of_soft_distance_correlation},
the model performance peaks at $\epsilon = 0.5$, where half of $h$ dimensions are relaxed from the Distance Correlation constraint, and preference embeddings are still able to grow diversely in the remaining half dimensions.
As $\epsilon$ approaches 0, the Distance Correlation constraint becomes too loose to encourage the diversity of preferences, leading to a dramatically decreased performance.

\subsubsection{Effects of Cross-System Contrastive Learning}
\label{sec:experiments:cross_system_learning}

\begin{table*}[!ht]
\caption{Benchmark results on Movie-KG-Dataset.
Numbers \underline{underlined} are existing state-of-the-art performance, while best performance of the proposed models is marked in \textbf{bold}.
The average of 3 runs is reported to mitigate experimental randomness.
CKMPN is better than best baseline results with statistical significance $p<0.05$.
}
\label{tab:compare_baselines_movies}
\vspace{-0.3cm}
\centering
\small
\begin{tabular}{llllllllll}
\toprule
                                  & \multicolumn{3}{c}{Recall} & \multicolumn{3}{c}{ndcg}   & \multicolumn{3}{c}{Hit Ratio}  \\
~                                 & @20   & @60   & @100    & @20   & @60     & @100  & @20   & @60   & @100        \\
\midrule
BPRMF                                & 0.1387 &	0.1944&	0.2206&	0.0961&	0.1137&	0.1192&	0.1980&	0.2785&	0.3236
          \\
CKE & 0.1369&	0.1898&	0.2150&	0.0940&	0.1108&	0.1160&	0.1950&	0.2707&	0.3155
\\
KGAT&	\underline{0.1403}&	0.1928&	0.2185&	\underline{0.1006}&	0.1173&	0.1226&	0.1997&	0.2742&	0.3196
\\
KGIN&	0.1351&	\underline{0.2119}&	\underline{0.2445}&	0.0982&	\underline{0.1254}&	\underline{0.1322}&	\underline{0.2194}&	\underline{0.3081}&	\underline{0.3643}\\

\midrule
KMPN ($\epsilon=0.5, N_m=64$) & 0.1434&	0.2130&	0.2427&	0.1073&	0.1305&	0.1367&	0.2193&	0.3098&	0.3602
\\
NRMS-BERT& 0.1241&	0.1669&	0.1890&	0.1034&	0.1213&	0.1257&	0.1728&	0.2369&	0.2773
\\
CKMPN ($\lambda_{CS}=0.01)$ & \textbf{0.1457}&	\textbf{0.2157}&	\textbf{0.2462}&	\textbf{0.1149}&	\textbf{0.1417}&	\textbf{0.1482}&	\textbf{0.2266}&	\textbf{0.3153}&	\textbf{0.3668}
\\
\midrule
CKMPN (on the cold-start set)  & 0.1024&	0.1741&	0.2130&	0.0570&	0.0729&	0.0808&	0.1812&	0.2839&	0.3380
\\

\bottomrule
\end{tabular}
\vspace{-0.2cm}
\end{table*}

\begin{figure}
\centering     
\subfigure[Recall@20 (blue) and ndcg @20 (red)
]{\label{fig:effects_of_cs_loss_20}\includegraphics[width=0.49\linewidth]{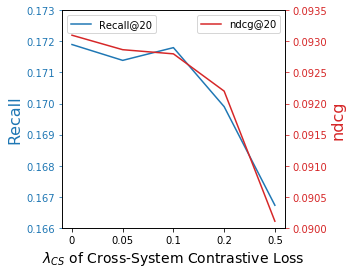}}
\hfill
\subfigure[Recall@100 (blue) and ndcg @100 (red)]{\label{fig:effects_of_cs_loss_100}\includegraphics[width=0.49\linewidth]{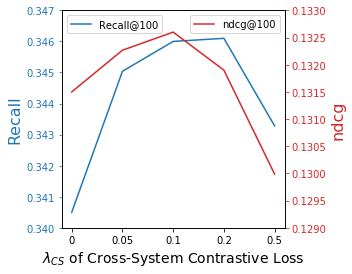}}
\vspace{-0.5cm}
\caption{System performance against the loss weight $\lambda_{CS}$ of Cross-System Contrastive Loss. The performance at top-$100$ (right) is greatly improved while keeping the metrics of top-$20$ (left) steady until $\lambda_{CS}>0.2$.}
\label{fig:effects_of_cs_loss}
\vspace{-0.3cm}
\end{figure}

We experimented with different $\lambda_{CS}$ in order to find the best configuration for the Cross-System Contrastive Loss.
As shown in Fig.~\ref{fig:effects_of_cs_loss_20}, the system performance of top-$20$ does not drop much for $\lambda_{CS}\leq0.2$ whereas Fig.~\ref{fig:effects_of_cs_loss_100} demonstrates that the performance at top-$100$ increases dramatically for $\lambda_{CS}\leq 0.2$ relative to a system without Cross-System Contrastive Loss.
This suggests that by incorporating Cross-System CL in our training, CKMPN is more capable of finding relevant items for users.






\vspace{-0.1cm}
\subsection{Benchmarking on Movie-KG-Dataset}
\label{sec:movie_experiments}

After verifying the effectiveness of our proposed models in Amazon-Book-Extended, we report their best performance on the newly collected movie dataset (Standard version with 19 days).\footnote{Extended version (39 days) will be released upon publication for larger applications.}
The performance of the same baseline systems is provided for inclusion (Table ~\ref{tab:compare_baselines_movies}).\footnote{The model performance on the cold-start test set is provided in Appendix A.}
The same performance boost is observed in KMPN relative to baselines.
For example, KMPN achieved 0.1434 Recall@20 and 0.1073 ndcg@20, which is higher than 0.1403 Recall@20 and 0.1006 ndcg@20 of the baselines.
CKMPN also achieved the best performance by incorporating content-based features from NRMS-BERT.
It outperforms KMPN in all metrics, showing a significant improvement in ndcg@100 (from 0.1367 to 0.1482) and Hit Ratio@100 (from 0.3602 to 0.3668) in particular.
Therefore, we can conclude that our method is applicable in multiple different datasets.

\textbf{Case Study:} An example output is presented in Table~\ref{tab:case_study}.
$\checkmark$/$\times$ indicates whether or not the movie appears in the top-$100$ recommendation list of the four models (KMPN/NRMS-BERT/Mixture of Expert/CKMPN).
This user has browsed Tenet (2020) directed by Christopher Nolan.
The movie Source Code (2011) and Tenet are both about time travel, but they have quite different film crews.
As a result, Source Code was considered positive by NRMS-BERT which evaluates on the movie description, but was considered negative by KG-based KMPN.
Combining the scores of two systems, MoE did not recommend the movie.
However, CKMPN complemented the failure of KMPN and gave a high score for this movie, by learning an content-aware item representation based on the representation of NRMS-BERT through Cross-System CL.
In contrast, Dunkirk (2017) is about war and history which is not in the same topic as Tenet.
But since they were directed by the same director, KMPN and CKMPN both recommended this movie, while MoE's prediction was negatively affected by NRMS-BERT.
This case study suggests that our Cross-System CL approach is an effective in-depth collaboration of two systems, outperforming the direct mixture of KMPN and NRMS-BERT.

\begin{table}[!h]
    \vspace{-0.2cm}
    \caption{Case study for a user who have browsed the movie Tenet (2020).
    Source Code (2011) has a similar genre, while Dunkirk (2017) has the same director.
    $\checkmark$/$\times$: whether or not the movie appears in the top-$100$ recommendation list of the models. MoE: Mixture of Expert.}
    \vspace{-0.3cm}
    \centering
    \begin{tabular}{ccccc}
    \toprule
        Item & KMPN & NRMS-BERT & MoE & CKMPN \\
        \midrule
        Source Code (2011) & $\times$ & $\checkmark$ & $\times$ & $\checkmark$
        \\
        Dunkirk (2017) &  $\checkmark$ & $\times$ & $\times$ & $\checkmark$
        \\
        \bottomrule
    \end{tabular}
    \vspace{-0.25cm}
    \label{tab:case_study}
\end{table}

We also present the model performance on the cold-start test set of the Movie-KG-dataset.
We recall that the cold-start set was created to evaluate model performance on users outside the training dataset.
Models are evaluated by predicting users' last 20\% viewed items with their first 80\% interactions included as history, under an extreme cold start situation where those users are completely unseen in the training.
As shown in the last section of Table~\ref{tab:compare_baselines_movies},
our best model CKMPN still achieved good performance for unseen users on all metrics, e.g., 0.1024 on Recall@20 and 0.3380 on Hit Ratio@100.
Compared with the standard test set, the performance did not deteriorated much, showing that our model still functions in the cold-start setting.

\subsection{Further Discussions}

In general, our proposed CKMPN has achieved substantial improvements on both datasets, especially on top-$60/100$ metrics. Industrial recommender systems usually follow a 2-step pipeline where a relatively large amount of items $K=60,100$ are firstly recalled by a Recall Model and then a Ranking Model is adopted to refine the list ranking.
This improvement presents more relevant items in the relatively coarser Recall output, which is appealing to industrial applications.
Also, CKMPN is much more preferred than the Mixture of Expert model in industrial applications, since it still produces independent user/item representations.
This feature enables fast and efficient match of users and items in hidden space with $O(log(n))$ query time complexity \cite{malkov2018efficient}.

From all these results we can conclude that by contrastive learning of cross-system user/item representations with pre-trained CBF features serving as proxies, CKMPN successfully fuses the external content-based information from NRMS-BERT.
It is also worth noting that this contrastive learning loss is not only limited to fusing textual modality, but can be applied to any other modalities that provide already good item/user representations for recommendation. 
We leave these promising extensions as future work.

\section{Conclusion}

We demonstrate that large Transformer models can be used to achieve content-sensitive collaborative filtering with the proposed fusion framework combining CF and CBF, two distinct paradigms.
To demonstrate the advantage of combining CF and CBF systems, two new datasets are presented and shared with the research community to facilitate the research of content-aware collaborative filtering:
(1) Amazon-Book-Extended which inherits a popular book recommendation dataset and has been newly extended with summary descriptions; 
(2) a new large-scale movie recommendation dataset, Movie-KG-Dataset, based on recently collected user interactions on a widely-used commercial web browser, accompanied by a knowledge graph and summary texts for the movies.

We present KMPN, a powerful CF model that utilizes \textit{Soft Distance Correlation Loss} and \textit{Reciprocal Ratio Negative Sampling} to outperform existing baseline models.
And we further show the effectiveness of a novel approach \textit{Cross-System Contrastive Learning}, which fuses the content-based features of a CBF model (NRMS-BERT) with KMPN to achieve a substantial improvement relative to models in literature.




\bibliographystyle{ACM-Reference-Format}
\bibliography{references}

\end{document}